**Title 1 (old):** A systematic framework for spatiotemporal modelling of COVID-19 disease
**Title 2:** An unbiased spatiotemporal risk model for COVID-19 with epidemiologically meaningful dynamics

**Author line:** Michał Paweł Michalak[1*], Jack Cordes[2], Agnieszka Kulawik[3], Sławomir Sitek[4], Sławomir Pytel[4], Elżbieta Zuzańska-Żyśko[4], Radosław Wieczorek[5]

**Author Affiliations:**

[1*] Institute of Earth Sciences, Faculty of Natural Sciences, University of Silesia in Katowice, Będzińska 60, 41-205 Sosnowiec, Poland, mimichalak@us.edu.pl

[2] Department of Epidemiology, Harvard T.H. Chan School of Public Health, 677 Huntington Avenue, Boston 02115, MA, USA

[3] Faculty of Science and Technology, University of Silesia in Katowice, Bankowa 14, 40-007 Katowice, Poland

[4] Institute of Social and Economic Geography and Spatial Management, Faculty of Natural Sciences, University of Silesia in Katowice, Będzińska 60, 41-205 Sosnowiec, Poland

[5] Institute of Mathematics, Faculty of Science and Technology, University of Silesia in Katowice, Bankowa 14, 40-007 Katowice, Poland

Word count: 4731

**Summary**

Spatiotemporal modelling of infectious diseases such as COVID-19 involves using a variety of epidemiological metrics such as regional proportion of cases or regional positivity rates. Although observing their changes over time is critical to estimate the regional disease burden, the dynamical properties of these measures as well as cross-relationships are not systematically explained. Here we provide a spatiotemporal framework composed of six commonly used and newly constructed epidemiological metrics and conduct a case study evaluation. We introduce a refined risk model that is biased neither by the differences in population sizes nor by the spatial heterogeneity of testing. In particular, the proposed methodology is useful for the unbiased identification of time periods with elevated COVID-19 risk, without sensitivity to spatial heterogeneity of neither population nor testing. We offer a case study in Poland that shows improvement over the bias of currently used methods and we believe our method can be implemented in other areas of the world. Our results also provide insights regarding regional prioritization of testing and the consequences of potential synchronization of epidemics between regions.

**Main Text**

**Introduction**

COVID-19 disease is caused by the novel coronavirus SARS-CoV-2 which was first discovered in late 2019 in Wuhan, Hubei Province, China. The main signs of infection include respiratory symptoms, fever, cough and breathing difficulties [1]. Spatiotemporal analysis plays a critical role in estimating the disease burden in specific regions [2], and it has been studied in many countries, including China [3,4], Spain [5], Italy [6,7], Sweden [8], Israel [9], Brazil [10] and United States [11,12]. It is important to note that comparisons between regions may be challenging, and this is not only because of differences in population sizes but also differences in health policies (e.g. testing regimes) that can change over time [11]. For example, one of the most common limitations raised when using spatiotemporal approaches relates to the lack of incorporation of the spatial heterogeneity of testing [13–15]. This omission may be misleading to public health officials in terms of an adequate public health response in regions with relatively high or low testing capabilities. From among many measures applicable in the spatiotemporal modelling of the COVID-19 disease, the following attracted the greatest attention: local [6] and global [16] cumulative number of cases, different versions of population-based relative risk (observed cases/expected cases) [14,17], testing rates (tests/population) [18], local [18] and global [16] positivity rates (confirmed cases/tests) [18,19] and population-based positivity (confirmed cases/population) [15,19]. Although potentially useful, the above measures were not investigated for their suitability in dynamical modelling of infectious disease. In particular, to the best of our knowledge, their dynamical properties and cross-relationships were not systematically explained. However, this has crucial importance for guiding the public health response: for example, it may be decided that an uninterrupted increase in cumulative version of population-based relative risk [14] entails the introduction of specific non-pharmaceutical interventions (NPIs) [20] in specific regions. Moreover, from the definition of these measures it follows that all of them are sensitive to spatial heterogeneity of either population or testing or both in their identification of elevated COVID-19, leading to bias.

This study is intended to offer a systematic contribution regarding a variety of spatiotemporal epidemiological measures independently used in both comparative [14,18,19] and non-comparative [16] contexts. We reveal dynamical properties and prove relationships between the commonly used and newly constructed epidemiological metrics with a case study application. The proposed methodology has the potential to enhance the framework of infectious disease modelling and provides insights into how a more harmonized management of the crisis can be achieved.

**Simultaneous standardisation with respect to population and testing**

The notion of relative risk is often used to investigate the spatial distribution of cases [17,21,22] and is inherently related to the concept of *indirect standardisation* [22]. It involves calculating the standardised incidence ratio (SIR) which accounts for the differences in population sizes among regions. Typically, this value depends on individual daily infection rates and is calculated as the ratio between the observed number of cases (infections) in a region and the expected number of cases based on the regional population. Values greater than one suggest an elevated risk compared to the population average which may indicate infection clusters or a greater number of vulnerable groups [17]. To alleviate the impact of daily fluctuations and create a framework for comparing the present state with historical reference, a cumulative model can be applied [14], which we refer to as the cumulative standardised incidence ratio (CSIR). It is calculated for a specific day $t$ as the ratio of the confirmed number of cases since the outbreak of the pandemic, including day $t$, to the expected cumulative number of cases on day $t$. We mathematically show that from a dynamical viewpoint it is equivalent to the regional cumulative proportion of cases; that is, for a specific day $t$ and region $i$, it may decrease or increase depending on whether the proportion of cases for $i$ on day $t$ is lower or greater than the cumulative proportion for $i$ on day $t-1$. The limitation of using this risk estimate to compare regions is that it is only unbiased if there is spatial homogeneity regarding testing intensity. To overcome this problem, we first applied an analogous procedure for the data related to regional testing. Namely, we calculated a quantity that we refer to as the cumulative standardised testability ratio (CSTR). It is calculated as the proportion of observed tests to expected tests in a given region. It is important to note that the resulting value can be interpreted as an estimate of the relative safety. This is because the greater the quantity is, the more efficiently NPIs can be applied. We then divided the CSIR by the CSTR to get a refined estimate of the relative risk, which we call the weighted cumulative standardized incidence ratio (WCSIR). It follows that the WCSIR remains unchanged from the CSIR only if the CSTR is equal to one, that is, the number of tests is equal to the expected number of tests. Otherwise, the relative risk increases or decreases depending on whether the CSTR is smaller than or greater than one, respectively. The resulting risk estimate is therefore biased neither by differences in population sizes nor by differences in amount of testing in specific regions. In other words, the WCSIR measure allows testing intensity to be heterogeneous and it captures the change in relative risk by honouring the following expectations:

1) For a specific region, a risk measure should decrease/increase if the regional infection rate (regional infected/global infected) decreases/increases and the analogous test rate (regional tests/global tests) increases/decreases.
2) For a specific region, a risk measure should decrease/increase if the cumulative global positivity rate (GPR) (cumulative positive/cumulative tests) increases/decreases, while the analogous local positivity rate (LPR) decreases/increases.

Alternatively, but equivalently (Theorem 2, Supplementary Materials), the WCSIR could be conceptualized as the ratio of LPR to GPR for a given region. Therefore, WCSIR equals one only if LPR=GPR, otherwise the relative risk increases or decreases depending on whether the LPR is greater than or less than GPR, respectively. As such, we anticipate our study to be a starting point for considering more sophisticated models of relative risk. For example, even if the dynamics of the regional proportion of cases show a positive trend for a given region, our methodology can classify this region as one with decreasing risk if the local and global positivity ratios are in opposite directions (i.e. are decreasing and increasing respectively).

**Case study motivation**

On March 4th, 2020, the first confirmed case was registered in Poland [23]. Four days later, cases were identified in the densely populated Silesian region [24]. As of August 17, 2020, 57,286 cases were verified in Poland, with 18,874 cases (34.8%) attributed to the Silesian region [25]. The core of the Silesian region is referred to as the Katowice conurbation, a polycentric area consisting of 16 towns and approximately 2 million people with a population density of 1,485 per km$^2$ (24). The largest urban centre is Katowice with 280 thousand inhabitants [26]. The concentration of public health efforts in Silesia follows from its large proportion of the population employed in industry (28.7%), compared to the mean-country value of 20.6%. Of those employed in industry in the area, 16.7% are employed in mining and exploration, which is also the largest value in Poland (mean-country value of 4.6 %) [27].

After the first case in Poland was identified, NPIs were introduced, including cancelling mass events and closing borders, schools, and universities among other measures [28,29]. These interventions helped flatten the curve of total infected individuals and delayed the peak of the disease burden (Fig. S1). However, the lockdown measures applied were ultimately insufficient in terms of containing the spread of the disease in the densely populated and relatively industrially oriented Silesian region. Because it is difficult to apply social distancing recommendations [1] in crowded mine shafts, it is hypothesized that the spread of COVID-19 in the Silesian region was facilitated by miners who might have acted as asymptomatic carriers. This has critical importance as it has been demonstrated that asymptomatic carriers play a vital role in the spread of the novel coronavirus [30] and undocumented infectious cases can facilitate rapid dissemination [31]. Indeed, according to partial results related to 50,053 screening tests within a group of mines conducted between May 7, 2020 and June 25, 2020 as a proxy for the whole mining population, nearly 98 percent of the infected mine employees did not indicate any symptoms [32,33]. As of August 11, 7,934 miners were tested positive for COVID-19 which yields nearly 44% of infections in Silesian region [34]. Although the time period related to screening tests in mines resulted in greater population-based relative risk for Silesia, two major problems remain: 1) whether the decision on screening tests was a result of an already deteriorating epidemiological situation in Silesia – with the beginning of this deterioration being unknown and 2) whether in other regions with relatively low testing capabilities the risk related to COVID-19 disease exists but is undetected.

We evaluated the proposed methodology on infection and testing rates that were provided by Ministry of Health in Poland: between March 4 and August 17 (infections) as well as between May 11 and August 17 (tests). The data were provided for 16 administrative regions of Poland (abbreviations presented in Fig. S2) that span the area of 312,696 km$^2$. Because the official testing rates were published on a weekly basis, to estimate the testing rates for individual days, we used interpolation. Other official reports supplemented by reports from news sources (as of March 26) were also used to perform interpolation for individual days between March 26 and May 11. Complete details regarding interpolation are available in the Methods.

**Methods**

**Materials**

We collected daily data on infections in Poland starting March 4, 2020 until August 17, 2020 from the Ministry of Health in Poland. From a technical viewpoint, the data are stored in a data frame in which columns represent days, while rows are administrative regions. A limitation of using this data is that reporting inaccuracies were not always adequately described. Although we applied 18 corrections, sometimes the dates corresponding to false positive cases or detected duplicates of confirmed cases were not provided. The inaccuracies and the corresponding metadata including dates of errors and applied corrections are summarised in Table S1.

The data on regional testing are available for the time period between March 26, 2020 and August 17, 2020. The reports from May 11, 2020 onward were provided by the Ministry of Health in Poland on a weekly basis in the form of cumulative number of tests conducted for every region. It should be noted, however, that these cumulative data do not include the full day of publishing but are restricted to a specific hour of the day. For example, cumulative regional data on testing published on August 17 cover the time period between March 4 and August 17, 1:00 p.m. We also used official fragmentary reports as well as an unofficial incomplete report for March 26, 2020 from a news source (Table S2). For days preceding March 26, 2020 no data on regional testing are publicly available, therefore we excluded this period from the analysis. There is concern about limited reliability of testing data for Kielce region for which about 241 000 tests were erroneously registered [54]. Although the correction was applied on August 8 [54], the historical data were not officially corrected. We therefore subtracted the superfluous number of tests evenly throughout the time period for the Kielce region. Population census data were obtained from the official repository Demographic Yearbook of Poland [55].

Estimating relative risk

To estimate the relative risk in regions of interest we used the concept of *indirect standardisation* [21,22]. The general formula for estimating the relative risk can be written as follows: $\frac{O}{E}$, where $O$ and $E$ denote the observed and the expected number of cases respectively. Given a specific region $i$, to obtain $E_i$ it is first necessary to calculate a global ratio $r = \frac{\sum_i O_i}{\sum_i P_i}$, where $\sum_i P_i$ denotes the total population. It is then straightforward to calculate $E_i$ as $P_i r$, with $P_i$ being the population in region $i$. For example, if the proportion of cases in Poland is 2%, then the expected number of cases in Silesia would be 2% of the population of Silesia, which assumes a spatially homogenous distribution of cases [21]. We must now differentiate between two versions of calculating the relative risks in our study: Standardised Incidence Ratio (SIR), Cumulative Standardised Incidence Ratio (CSIR) and Weighted Cumulative Standardised Incidence Ratio (WCSIR) based on Cumulative Standardised Testability Ratio (CSTR).

For the SIR, the totals of observed number of cases and the expected number of cases $E_i$ refer to the daily number of cases. If we denote $x_{i,t}$ as the observed number of cases on day $t$, and $E_{i,t}$ as the expected number of cases on day $t$, we can write the following formula:

$$SIR_i(t) = \frac{x_{i,t}}{E_{i,t}} = \frac{x_{i,t}}{P_i r_t} = \frac{x_{i,t}}{P_i \frac{\sum_i x_{i,t}}{\sum_i P_i}} \qquad (1)$$

For the CSIR, we assume that, for a specific day $t$, the observed number of cases in region $i$ is the sum of the observed number of cases in $i$ for days $j \in [1, t]$. Similarly, to calculate the corresponding ratio we assume that the observed number of cases is the sum of all cases that were confirmed in Poland for days $j \in [1, t]$. Then, the ratio is multiplied by the population size in $i$ to get the expected number of cases. The CSIR is then calculated as the proportion between the observed and expected number of cases. If we denote $O_{i,t-1} = \sum_{j=1}^{t-1} x_{i,j}$ as the cumulative number of cases for region $i$ up to day $t - 1$, and $E_{i,1:t}$ as the expected cumulative number of cases for $i$ on day $t$, then the formula for CSIR can be written as follows:

$$CSIR_i(t) = \frac{O_{i,t-1} + x_{i,t}}{E_{i,1:t}} = \frac{O_{i,t-1} + x_{i,t}}{P_i \frac{\sum_i (O_{i,t-1} + x_{i,t})}{\sum_i P_i}} \qquad (2)$$

From this it follows that the current value of CSIR is to a large extent governed by the past (cumulative cases from days 1 up to $t-1$) and the contribution of present day $t$ weakens with time. This explains why the curves are smooth: the present has a smaller effect than the past. To sum up, the past contributes primarily to the present state and, with time, its role in shaping the present increases.

Similarly, the information on the number of tests conducted for each region can be employed to estimate the relative safety (not to be confused with the safety perspective presented in Fig. S11). The corresponding Cumulative Standardised Testability Ratio (CSTR) was calculated in an analogous procedure to that of the CSIR. If we denote $T_{i,t-1}$ as the cumulative number of tests conducted for region $i$ up to day $t-1$, $y_{i,t}$ as the number of tests conducted for $i$ on day $t$ and $TE_{i,1:t}$ as the cumulative expected number of tests for $i$ on day $t$, then the formula for CSTR can be written as follows:

$$CSTR_i(t) = \frac{T_{i,t-1}+y_{i,t}}{TE_{i,1:t}} = \frac{T_{i,t-1}+y_{i,t}}{P_i \frac{\Sigma_i(T_{i,t-1}+y_{i,t})}{\Sigma_i P_i}} \qquad (3)$$

The interpolation procedure was conducted as follows. We first used `interpolate_tests` function to estimate the cumulative number of tests for individual days assuming a constant intercept: for example, if the cumulative number of tests for region $i$ were 1000 and 8000 on May 11 and May 18 respectively, then the intercept for every day related to this time window is equal to $\frac{8000-1000}{7} = \frac{7000}{7} = 1000$. Then, to obtain the approximate value of CSTRs for individual days, `relrisk` function was used. We note that the assumption of constant intercept may not be realistic but we are not ready to commit to the idea of the best interpolation method in case of differences in temporal resolution between data on infections and testing. To avoid edge-effects that are inherent for the interpolation method, statistical approaches including best-fitting curves could be employed. However, the main disadvantage of the statistical procedure is that one cannot expect that the official data of cumulative number of tests will be honoured at nodes.

The interpolation enabled the weighted model to be obtained through the division of data frames corresponding to data on infections and testing. The corresponding equation of the resulting WCSIR is as follows: $(Observed\ Cases/Expected\ Cases)/(Observed\ Tests/Expected\ Tests)$.

$$WCSIR_i(t) = \frac{\frac{O_{i,t-1}+x_{i,t}}{E_{i,1:t}}}{\frac{T_{i,t-1}+y_{i,t}}{TE_{i,1:t}}} = \frac{\frac{O_{i,t-1}+x_{i,t}}{P_i\frac{\Sigma_i(O_{i,t-1}+x_{i,t})}{\Sigma_i P_i}}}{\frac{T_{i,t-1}+y_{i,t}}{P_i\frac{\Sigma_i(T_{i,t-1}+y_{i,t})}{\Sigma_i P_i}}} \qquad (4)$$

From a methodological viewpoint, we note that the curves presented here could also be viewed as a measure of relative safety in the form of $\frac{1}{relative\ risk}$. This safety perspective provides a better visualization of regions with lowest values of relative risk that in the CSIR or WCSIR models are difficult to distinguish (Fig. S11).

Local and global positivity ratios (LPR and GPR)

Because the relationship between local and global dynamics of the cumulative proportion of positive cases exerts influence on the dynamics of WCSIR, we provide formulas for these measures:

$$LPR_i(t) = \frac{O_{i,t}}{T_{i,t}} \qquad (5), \qquad GPR(t) = \frac{O(t)}{T(t)} \qquad (6)$$

The computational objectives corresponding to functions included in the computer code are summarised in Table 2. We used the following R packages: `dplyr` [56], `ggplot2` [57], `ggpubr` [58], `reshape2` [59], `tibble` [60], `sf` [61], `tmap` [62], `broom` [63], `plotly` [64] and `magrittr` [65].

**Results**

**Unweighted risk**

The SIR analysis reveals that in the Silesian region (12 KAT) the relative risk values were largely greater than one since mid-April and on five days in late April and early May this value was greater than 3, all before the decision to screen for COVID-19 in mines (Fig. 1A). In the 28 days after implementation of this testing policy, SIR values were always greater than 3, with maximum value 6.92 on May 12th and fluctuations between 4 and 6. However, greater than expected number of cases since mid-April suggest that the outbreak might have originated earlier than the decision to test miners. Indeed, CSIR curves show a rising trend prior to the decision to implement screening tests in mines in Silesia largely after the Easter holiday (Fig. 1B). The nearly monotonously rising trajectory of both CSIR between mid-April and mid-June denotes the progressive relative deterioration of the epidemiological situation in Silesia for this time period.

**Weighted risk**

However, as discussed above, the estimates of relative risks presented in Fig. 1 are biased by the regional differences in testing intensity. Figure S3 shows that throughout the epidemic the highest testing intensity was observed for the Warszawa region (7 WAR), with lowest positions occupied by the Opole (8 OPO) and Rzeszów (9 RZE) regions.

According to the WCSIR model (Fig. 2; Fig. 3 – lowest part) the Opole region (8 OPO) was the least safe region during the height of the epidemic in Poland (11 April-18 May), and at the end of the study period occupies second position, with a significant distance to the third ranked Rzeszów region (9 RZE). Although positive differences between CSIR and WCSIR for both Silesia (12 KAT) and Opole (8 OPO) can be observed (Figs. 2, 3), for the Silesia region this difference is smaller and the corresponding curves are more or less parallel (Fig. 2). This is not the case for the Opole region, whose CSIR and WCSIR curves show a greater difference and a diverging pattern in mid-May: while the WCSIR was increasing during May 13-18 with a simultaneous decrease in CSTR, an increase in CSIR was observed only between May 15-17. This means that the weighted risk estimate may increase even if the unweighted risk decreases.

**Dynamical patterns**

A more comprehensive analysis of the relationship between cumulative measures (CSTR, CSIR, WCSIR and the LPR-GPR tandem (Fig. S4)) can be conducted by plotting daily values of two selected measures on a Cartesian plane (Figs. 4, S5-S10). In summary of all sixteen regions, we highlighted six potentially distinct trajectories that emerged. Fig. 4A shows the ideal situation when CSIR and WCSIR decrease together throughout the period, as exemplified by the Wrocław region (1 WRO). This pattern can be summarised that although sometimes CSTR is increasing, the CSIR is decreasing (Fig. S5A). The second pattern for Warszawa (7 WAR) region shows that the decreasing value of CSIR is associated with an increasing value of WCSIR and LPR (Figs. 4B, S6B). The trajectory of the Opole region (8 OPO) region represents the third pattern: while the CSIR was approximately constant, the WCSIR was rising fast (Fig. 4C), confirming the divergent pattern in mid-May observed in Figure 2. We note however that the LPR and WCSIR show a decreasing trend since June (Figs. 4C, S6C) in Opole region. Figure 4D shows the fourth pattern with an undesireable change of the trajectory of the relationship, as typified by the Rzeszów region (9 RZE) since mid-June. For Silesia (12 KAT) (Fig. 4E) we observe the fifth pattern of a plateau in the relationship after initial growth of CSIR and WCSIR. This may be due to largely decreasing LPR since June when it reached a maximum value of 10.25% (Fig. S4). The last pattern, exemplified by the Poznań (15 POZ) region (Figs. S5F, Figs. S8-S9F), shows a change in direction of CSTR trajectory that started to decrease in mid-June. Although at the beginning of this change LPR was still decreasing, within a month from this change it started to increase (Fig. S10F) with undulating CSIR and WCSIR (Figs. 4F, S5F).

**Discussion**

In this article, we highlighted the properties and relationships between commonly used and newly constructed epidemiological measures applicable in spatiotemporal infectious disease relative risk modelling. We stressed the role of including information on testing intensity in estimating the relative risks of COVID-19 infection, with a case study in Poland. The weighted approach is particularly useful when spatial homogeneity in testing intensity cannot be assumed, which was the case in our example. For instance, as of August 17, the Warszawa region was tested nearly 4.5 times more intensely (CSTR=1.66) than the Rzeszów region (CSTR=0.37). Given these disproportions, inferring the epidemic dynamics from the confirmed number of cases is not justified [35]. We also show the official statements regarding the Silesia region as the most tested (CSTR never greater than 1, as of August 17) [36], the epidemiologically safest region [37], or epidemiologically unexceptional (both CSIR and WCSIR>1) [38] to be false.

The refined results could be utilized by authorities and health crisis managers to introduce more integrated NPI policies for adjacent regions that can be epidemiologically synchronized [39,40]. In our case, the similarity between WCSIR values and relatively high positivity rates throughout the study period suggest that this synchronization may be the case for the Opole and Silesia regions. As of now in these regions, differences in organising public gatherings remain: for example in the Opole region, church authorities allowed the organisation of city-wide processions at the Feast of Corpus Christi (June 11) [41], whereas in Silesia mass gatherings of this kind were forbidden [42]. Because public gatherings played a vital role in the spread of the 1918 influenza pandemic [20,43], it is now necessary to stress the significance of the joint effect of testing and infection rates to prevent downplaying the epidemiological risk in poorly tested regions.

We moreover illustrated that incorporating the historical data for calculating the relative risk can be beneficial to better identify time periods related to investigated trends. Using a cumulative model of relative risk, we demonstrate that systematic growth in infections already started in mid-April, three weeks before the decision to implement screening tests in mines in Silesia. This is particularly concerning as the lack of screening tests at the time of potentially greater, yet unknown, mobility corresponding to Easter may have facilitated the local spread of the disease among these mostly asymptomatic and thus undetected carriers. We note however that the epidemiological deterioration did not affect all mines equally and it was highly variable throughout the study period: while at the end of June the positivity ratio calculated for a group of several mines did not exceed 4% (1,862 confirmed cases/50,053 tests) [33], one month later a screening test conducted for one mine from this group revealed about 35% population-based positivity (156 confirmed cases/452 employees) [44]. Surprisingly, the systematic deterioration and highest positivity ratio in May for Silesia coincides with a temporal concentration of increase in one particular mine in which the population-based positivity increased from about 5% (245/4,982 employees, as of May 14) to 28% (as of May 24), to 33% (as of July 29).

The investigation of COVID-19 spread in similar conditions was already carried out in the densely populated region of Buenos Aires with 13 million inhabitants in 41 districts [45]. This research was based on analysing anonymized mobile phones and it revealed that the spread of the disease was radial in nature: from the central city of Buenos Aires, through the suburban districts, to the neighbouring regions. There are however two major differences underlying the spread of the disease in Buenos Aires versus the Silesian region. While the first difference is related to a very specific spatial structure of the Silesian region, the second points at the greater role of industry rather than population density in the spread of the disease. Other studies have likewise shown that population density [15,46], age structure [15], and socioeconomic status [11,19,47] among other variables affect COVID-19 transmission patterns, but this study adds knowledge regarding how the area's relative dominance in an economic sector can play a role in the transmission. We note however that the mining industry should be only regarded as a proxy of the infectious potential of large industrial plants and indeed similar events were registered in other regions (e.g. in 15 POZ region in meat-processing company in early August – 313 confirmed cases/800 employees) [48]. Given these plants may attract employees from more distant localities, they have the potential to synchronize the epidemics at the sub-regional level, sometimes trespassing the administrative borders of a higher level [40]. Therefore, particular attention should be paid if this synchronization may be the case for administrative regions with different testing capabilities.

The main limitation of this study follows from the incompleteness of official data related to testing data that are publicly available only since May 11th (see details in Methods). For the time period between March 26 and May 11, the results of weighting approaches and positivity ratios are estimated with greater uncertainty and we urge more caution in interpreting estimates at the beginning of the time period. The confirmed yet unexplained underreporting for Silesia resulting in about an 8% (as of July 9) underestimate of cases as well as limited reliability of testing data for Kielce region [49] poses additional interpretation difficulties [50]. An additional limitation is that there may be other unknown differences in testing regimes influencing the results. For example, poorly tested regions may decide that only very suspicious cases will be tested which will result in a relatively high positivity ratio [16]. We also did not include information on recovered individuals, defined as those with two subsequent negative tests [51] which underestimates the positivity ratios for the time period in which the number of recovered increases. It should also be stressed that the theorem regarding the dynamics of WCSIR has the form of an implication, so one is not allowed to establish a causal relationship between a decrease of WCSIR and either of the alternatives in the implication given. We stress that the WCSIR can be thought of in two alternative ways (CSIR/CSTR or LPR/GPR, Observation 1 in Theorem 2). This flexibility affords opportunities overcome potential difficulties with obtaining population data, which may be the case if a testing site cannot be easily split into regions with known population.

Because testing capabilities are always limited, the obtained results after weighting could also be used to consider regional prioritization in the availability of tests. For example, from Table 1 it could be inferred that the following regions are in particular need for increasing the intensity of testing: Opole (8 OPO: + 170.53%) and Rzeszów (9 RZE: + 171.41%). A more meaningful analysis of testing capabilities should however always involve the temporal aspect (Fig. S3) and the relationship between the epidemiological indices that we proposed and developed throughout time (Figs. 4, S5-S10). Therefore, we support calls for the radical increase in the identification of positive cases and accompanying isolation as well as encourage behavioural changes and increase awareness of the disease to help reduce the spread of COVID-19 [31,52,53]. We also believe that this method holds promise to guide efforts in countries without a robust health care infrastructure and to counter a misinterpretation of the perception of a high relative risk in densely tested areas compared to areas with low apparent relative risk due to limited testing availability, when in fact a high risk of COVID-19 may simply be undetected.

**Table 1** Estimates of relative risk (as of August 17) using unweighted (CSIR) and weighted (WCSIR) approaches, relative changes and their corresponding positions

| Region | CSIR | Rank 1 | Relative Change (%) | WCSIR | Rank 2 |
|---|---|---|---|---|---|
| 1 WRO | 0.85 | **7** | -6.70 | 0.79 | **9** |
| 2 BYD | 0.33 | **15** | +11.60 | 0.37 | **16** |
| 3 LUB | 0.40 | **12** | +31.76 | 0.53 | **12** |
| 4 GOR | 0.39 | **13** | +112.31 | 0.83 | **8** |
| 5 LOD | 1.27 | **2** | -2.03 | 1.24 | **4** |
| 6 KRA | 1.09 | **3** | -0.73 | 1.09 | **5** |
| 7 WAR | 0.99 | **4** | -39.66 | 0.60 | **11** |
| 8 OPO | 0.98 | **5** | +170.53 | 2.66 | **2** |
| 9 RZE | 0.59 | **10** | +171.41 | 1.61 | **3** |
| 10 BIA | 0.65 | **8** | +19.97 | 0.78 | **10** |
| 11 GDA | 0.47 | **11** | -15.24 | 0.40 | **15** |
| 12 KAT | 2.80 | **1** | +3.10 | 2.89 | **1** |
| 13 KIE | 0.64 | **9** | +33.05 | 0.85 | **7** |
| 14 OLS | 0.30 | **16** | +43.36 | 0.42 | **14** |
| 15 POZ | 0.91 | **6** | -5.82 | 0.86 | **6** |
| 16 SZC | 0.39 | **14** | +13.04 | 0.44 | **13** |

**Table 2** The description of functions included in the computer code (R script)

| Function | What does it do? |
|---|---|
| cumulate_df | A function that cumulates a data frame |
| relrisk | Calculates SIR. Note that it can also be used to provide input to interpolate_tests function. |
| relrisk_cum | Calculates CSIR. |
| sum_cum | Calculates cumulative sum of observed cases for every day. |
| interpolate_tests | Calculates CSTR. |
| weighted_risk | Calculates WCSIR. |
| reorder | It is used to properly assign risk CSIR and WCSIR values to the polygons in shapefile. |

**Figures**

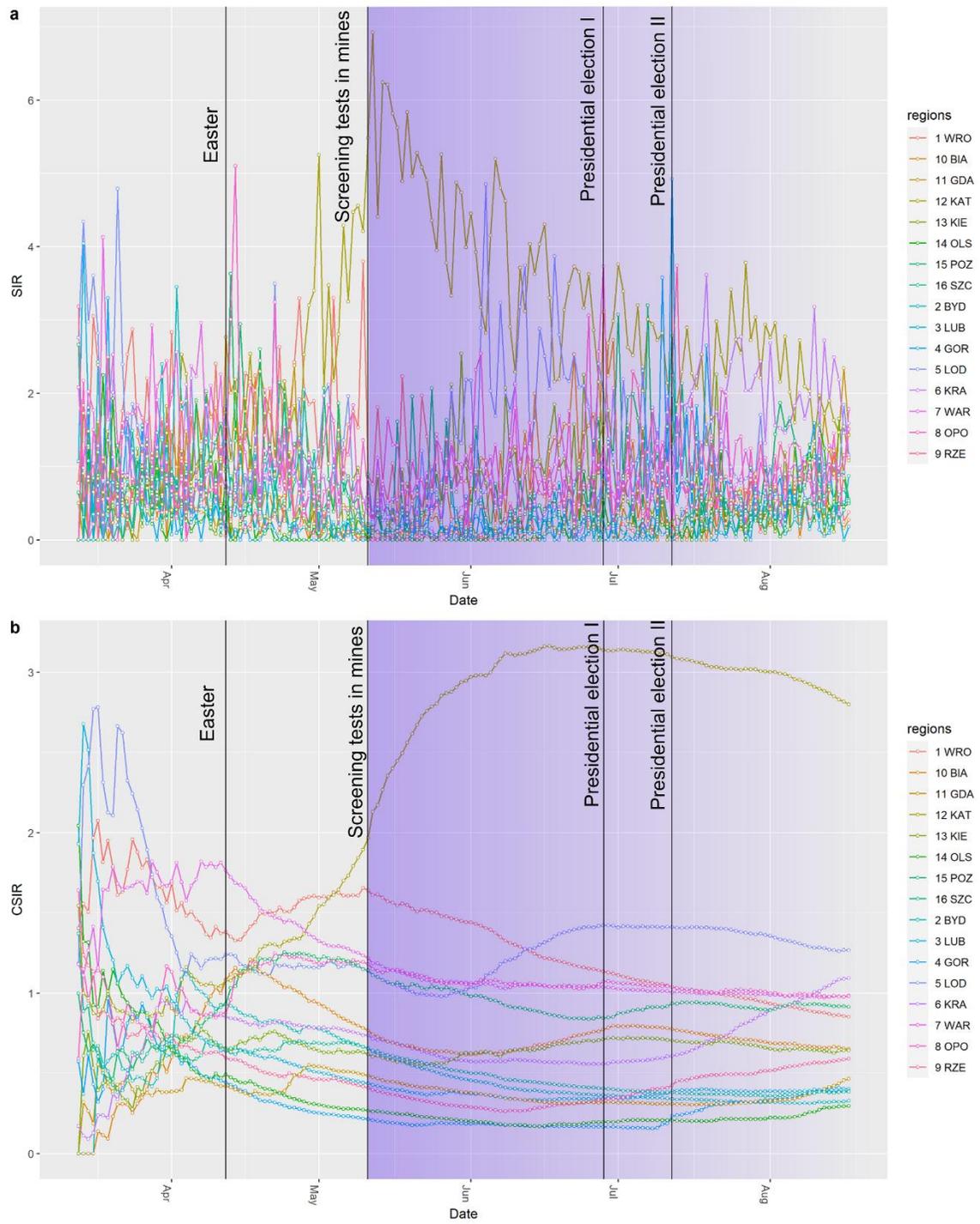

**Fig. 1** Daily time series of (A) SIR and (B) CSIR by region in Poland. The gradient denotes the assumed decreasing intensity of screening tests with time

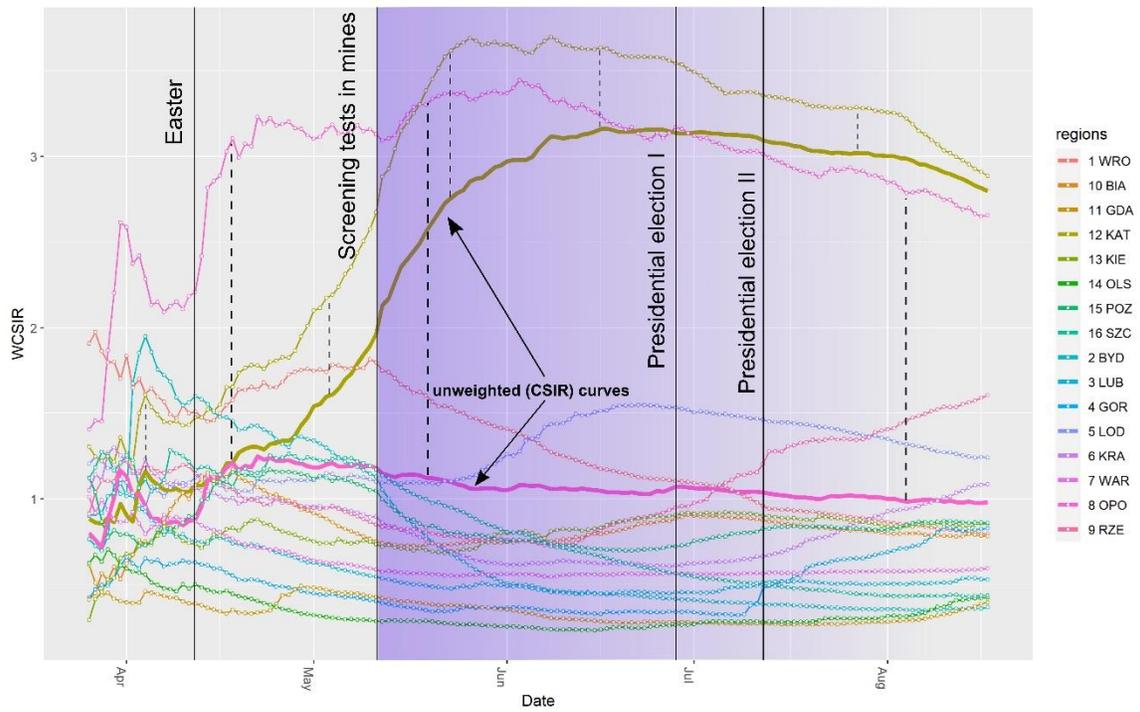

**Fig. 2** WCSIR by region in Poland. Note that we included two unweighted CSIR curves (bolded) to illustrate the impact of weighting. The gradient denotes the assumed decreasing intensity of screening tests with time

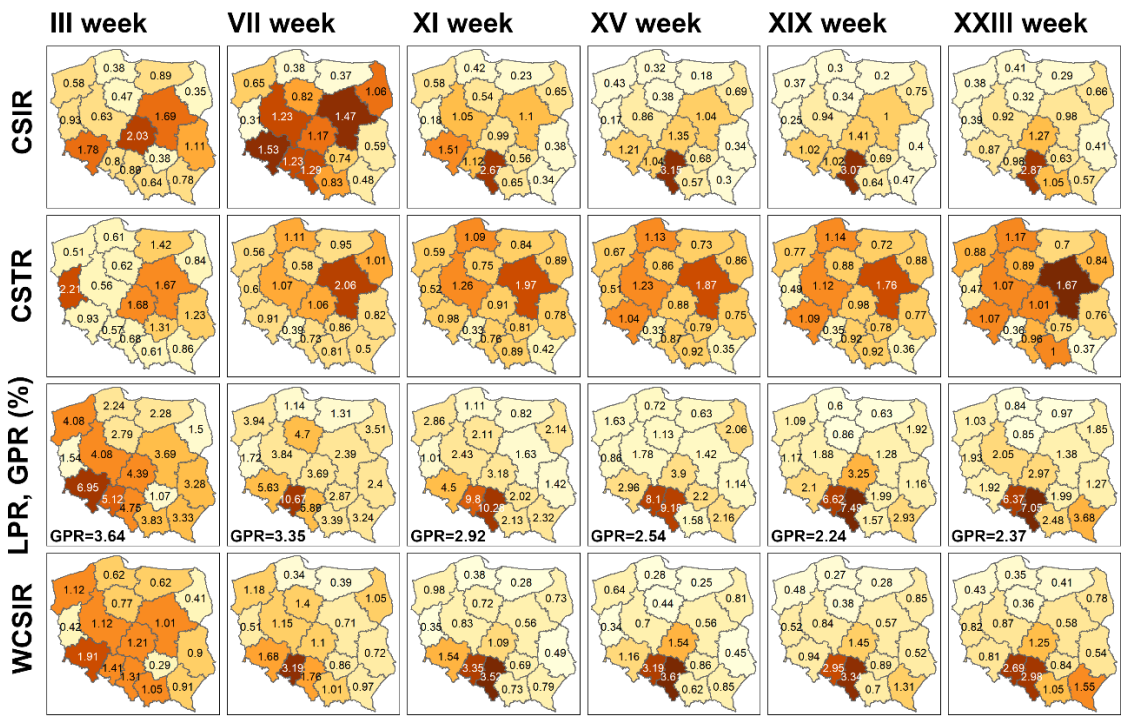

**Fig. 3** COVID-19 epidemiological measures for administrative regions in Poland. (CSIR – Cumulative Standardised Incidence Ratio, CSTR – Cumulative Standardised Testability Ratio, LPR – Local Positivity Rate, GPR – Global Positivity Rate, WCSIR – Weighted Cumulative Standardised Incidence Ratio)

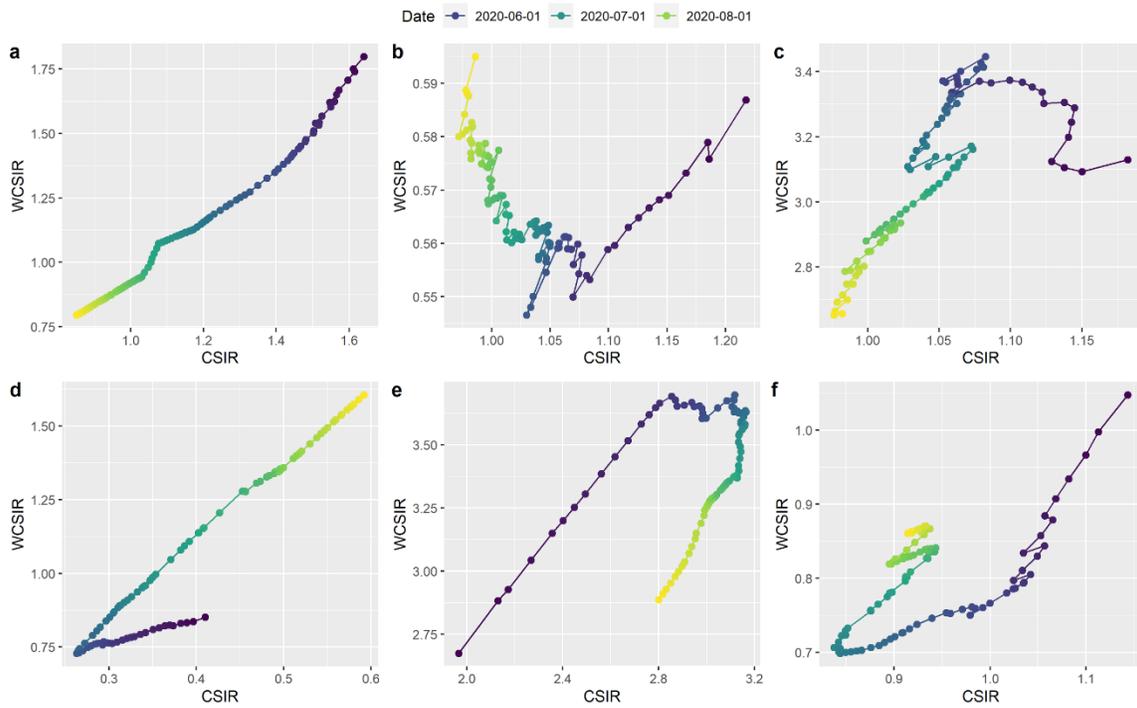

**Fig. 4** Relationship between CSIR and WCSIR curves for individual regions representing six distinct patterns: (A) 1 WRO: Both CSIR and WCSIR values are decreasing, (B) 7 WAR: CSIR is decreasing and WCSIR is increasing, (C) 8 OPO: divergent pattern of CSIR and WCSIR, (D) 9 RZE: both CSIR and WCSIR are increasing, (E) 12 KAT: proportional growth of CSIR and WCSIR, then the relationship plateaus, (F) 15 POZ: a change in trajectory followed by a zigzag pattern between CSIR and WCSIR

**Financial support**: The study (APC) was funded by University of Silesia in Katowice.

**Conflicts of Interest: None**

**Data Availability Statement**

Processed data on infections and testing are available at

https://github.com/michalmichalak997/COVID-19/blob/master/Data%20and%20code/Data%20and%20code_December_4_update.zip . Original data on infections and testing are available at https://twitter.com/mz_gov_pl?lang=pl. Processed data that are used for generating individual plots are available at

https://github.com/michalmichalak997/COVID-19/blob/master/Data%20and%20code/Data%20for%20figures_December_4_update.zip

Population census data are available at https://stat.gov.pl/en/ .

**Code availability**

Computer code is available at https://github.com/michalmichalak997/COVID-19/blob/master/README.md. Interactive plots are publicly available at

https://michalmichalak997.shinyapps.io/shiny_corona/

**Title 1 (old):** A systematic framework for spatiotemporal modelling of COVID-19 disease
**Title 2:** An unbiased spatiotemporal risk model for COVID-19 with epidemiologically meaningful dynamics

Authors: Michał Paweł Michalak, Jack Cordes, Agnieszka Kulawik, Sławomir Sitek, Sławomir Pytel, Elżbieta Zuzańska-Żyśko, Radosław Wieczorek

**'Supplementary Material'**

**Mathematical formalism**

**Theorem 1.** For a specific region $i$, the $CSIR$ on day $t$ will decrease if the proportion of new cases for this region on day $t$ will be not greater than the proportion of all cases for this region on day $t-1$.

*Proof.*

Let $i$ be a specific region and $CSIR_i(t-1) = O_{i,t-1}/E_{i,1:(t-1)}$ and $CSIR_i(t) = (O_{i,t-1} + x_{i,t})/E_{i,1:t}$ be the cumulative estimates of relative risks for region $i$ on days $t-1$ and $t$, respectively, where $O_{i,t-1}$ denotes the total (historical) sum of observed number of cases for $i$ registered on day $t-1$, $x_{i,t}$ denotes new cases for $i$ confirmed on day $t$, and $E_{i,1:(t-1)}$ and $E_{i,1:t}$ denote the expected cumulative number of cases for $i$ on days $t-1$ and $t$, respectively. The decrease in $CSIR$ between days $t-1$ and $t$ may be written in the form of an inequality:

$$\frac{O_{i,t-1}+x_{i,t}}{E_{i,1:t}} \leq \frac{O_{i,t-1}}{E_{i,1:(t-1)}} \leftrightarrow \qquad (7)$$

$$O_{i,t-1} + x_{i,t} \leq \frac{O_{i,t-1}E_{i,1:t}}{E_{i,1:(t-1)}} \leftrightarrow \qquad (8)$$

$$x_{i,t} \leq \frac{O_{i,t-1}E_{i,1:t}}{E_{i,1:(t-1)}} - \frac{O_{i,t-1}E_{i,1:(t-1)}}{E_{i,1:(t-1)}} = \frac{O_{i,t-1}(E_{i,1:t}-E_{i,1:(t-1)})}{E_{i,1:(t-1)}} \leftrightarrow \qquad (9)$$

$$\frac{x_{i,t}}{O_{i,t}} \leq \frac{E_{i,1:t}-E_{i,1:(t-1)}}{E_{i,1:(t-1)}} \leftrightarrow \qquad (10)$$

But $E_{i,1:(t-1)} = P_i r_{t-1}$ and $E_{i,1:t} = P_i r_t$ with $r_{t-1} = \frac{O(t-1)}{\sum_i P_i}$ and $r_t = \frac{O(t-1)+x(t)}{\sum_i P_i}$, where $O(t-1)$ denotes the total (historical) sum of observed number of cases for all regions on day $t-1$, $x(t) = \sum_i x_{i,t}$ denotes the sum of cases confirmed for all regions on day $t$, $P_i$ is the population size in $i$, and $\sum_i P_i$ is the country-wide population. Thus,

$$\frac{x_{i,t}}{O_{i,t-1}} \leq \frac{P_i(r_t-r_{t-1})}{P_i r_{t-1}} = \frac{r_t-r_{t-1}}{r_{t-1}} \qquad (11)$$

Substituting for $r_t$ and $r_{t-1}$:

$$\frac{x_{i,t}}{O_{i,t-1}} \leq \frac{\frac{O(t-1)+x(t)}{\Sigma_i P_i} - \frac{O(t-1)}{\Sigma_i P_i}}{\frac{O(t-1)}{\Sigma_i P_i}} = \frac{\frac{x(t)}{\Sigma_i P_i}}{\frac{O(t-1)}{\Sigma_i P_i}} = \frac{x(t)}{O(t-1)} \qquad (12)$$

But the above result is equivalent to:

$$\frac{x_{i,t}}{x(t)} \leq \frac{O_{i,t-1}}{O(t-1)} \qquad (13)$$

Note that $\frac{x_{i,t}}{x(t)}$ is the proportion of cases for region $i$ on day t, while $\frac{O_{i,t-1}}{O(t-1)}$ is the "cumulative" proportion of cases for region $i$ on day $t-1$, i.e. the total (historical sum) of confirmed cases for $i$ up to day $t-1$, divided by the total (historical sum) of confirmed cases in the country up to day $t-1$.

It is also possible to write a simplified version of the above proof. Let $O(t) = \Sigma_i O_{i,t}$, $P = \Sigma_i P_i$, $\rho_i = \frac{P_i}{P}$ and $x(t) = \Sigma_i x_{i,t}$. Then

$$CSIR_i(t) = \frac{O_{i,t}}{E_{i,t}} = \frac{O_{i,t}}{\frac{P_i O(t)}{P}} = \frac{1}{\rho_i} \frac{O_{i,t}}{O(t)}. \qquad (14)$$

Then,

$$CSIR_i(t-1) \geq CSIR_i(t) \leftrightarrow \qquad (15)$$

$$\frac{1}{\rho_i} \frac{O_{i,t-1}}{O(t-1)} \geq \frac{1}{\rho_i} \frac{O_{i,t}}{O(t)} \leftrightarrow \qquad (16)$$

$$\frac{O_{i,t-1}}{O(t-1)} \geq \frac{O_{i,t}}{O(t)} \leftrightarrow \qquad (17)$$

$$O_{i,t-1} O(t) \geq O_{i,t} O(t-1) \qquad (18)$$

$$O_{i,t-1}\big(O(t-1) + x(t)\big) \geq (O_{i,t-1} + x_{i,t}) O(t-1) \qquad (19)$$

$$O_{i,t-1} x(t) \geq x_{i,t} O(t-1) \qquad (20)$$

$$\frac{O_{i,t-1}}{O(t-1)} \geq \frac{x_{i,t}}{x(t)} \qquad (21)$$

**Theorem 2.** The WCSIR will decrease if either of the two following conditions in the form of conjuctions are met (we note that an analogous theorem can be formulated for an increasing WCSIR):

(i) The CSIR is decreasing (proportion of infected for region $i$ on day $t$ is not greater than the cumulative proportion of infected for region $i$ on day $t-1$) and the CSTR is increasing (proportion of tests in region $i$ on day $t$ is not smaller than the cumulative proportion of tests on day $t-1$.)

(ii) LPR is decreasing (the ratio of positive cases for region $i$ on day $t$ is not greater than the cumulative ratio of positive cases on day $t-1$) and GPR is increasing (the ratio of positive cases for the whole country on day $t$ is not smaller than the cumulative ratio for the whole country of positive cases on day $t-1$).

*Proof.*

(i)

If $CSIR_i(t) \leq CSIR_i(t-1)$ and $CSTR_i(t) \geq CSTR_i(t-1)$, then

$WCSIR_i(t) = \frac{CSIR_i(t)}{CSTR_i(t)} \leq \frac{CSIR_i(t-1)}{CSTR_i(t-1)} = WCSIR_i(t-1).$

(ii)

Let us first prove a following observation.

Observation 1. $WCSIR_i(t) = \frac{LPR_i(t)}{GPR(t)}$.

*Proof.*

$$WCSIR_i(t) = \frac{\frac{O_{i,t}}{E_{i,1:t}}}{\frac{T_{i,t}}{TE_{i,1:t}}} = \frac{\frac{O_{i,t}}{T_{i,t}}}{\frac{E_{i,1:t}}{TE_{i,1:t}}} = \frac{LPR_i(t)}{P_i \frac{O(t)}{\sum_i P_i}} = \frac{LPR_i(t)}{GPR(t)}.$$

It results that if $LPR_i(t) \leq LPR_i(t-1)$ and $GPR(t) \geq GPR(t-1)$, then

$$WCSIR_i(t) = \frac{LPR_i(t)}{GPR(t)} \leq \frac{LPR_i(t-1)}{GPR(t-1)} = WCSIR_i(t-1).$$

**Supplementary Figures**

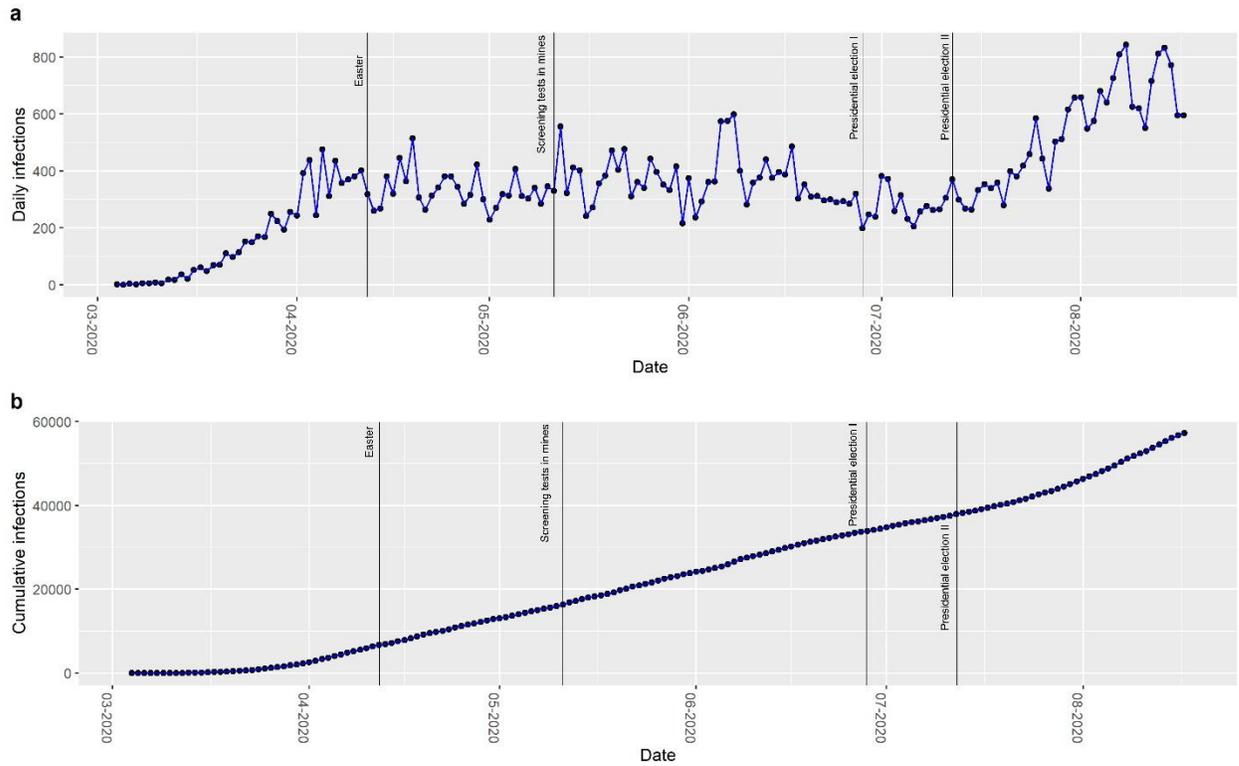

**Fig. S1** The development of the epidemic in Poland throughout the study period: (A) Number of daily infections; (B) Cumulative number of infections

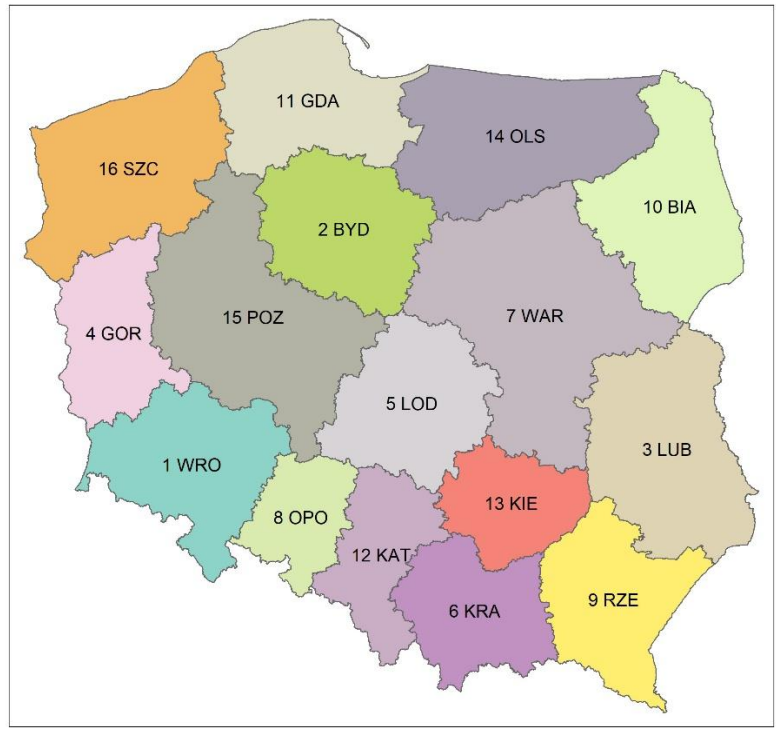

**Fig. S2** Administrative regions of Poland with full names and capitals (in parentheses): 1 WRO – dolnośląskie (Wrocław), 2 BYD – kujawsko-pomorskie (Bydgoszcz), 3 LUB – lubelskie (Lublin), 4 GOR – lubuskie (Gorzów Wielkopolski), 5 LOD – łódzkie (Łódź), 6 KRA – małopolskie (Kraków), 7 WAR – mazowieckie (Warszawa), 8 OPO – opolskie (Opole), 9 RZE – podkarpackie (Rzeszów), 10 BIA – podlaskie (Białystok), 11 GDA – pomorskie (Gdańsk) , 12 KAT – śląskie (Katowice), 13 KIE – świętokrzyskie (Kielce), 14 OLS – warmińsko-mazurskie (Olsztyn), 15 POZ – wielkopolskie (Poznań), 16 SZC – zachodniopomorskie (Szczecin)

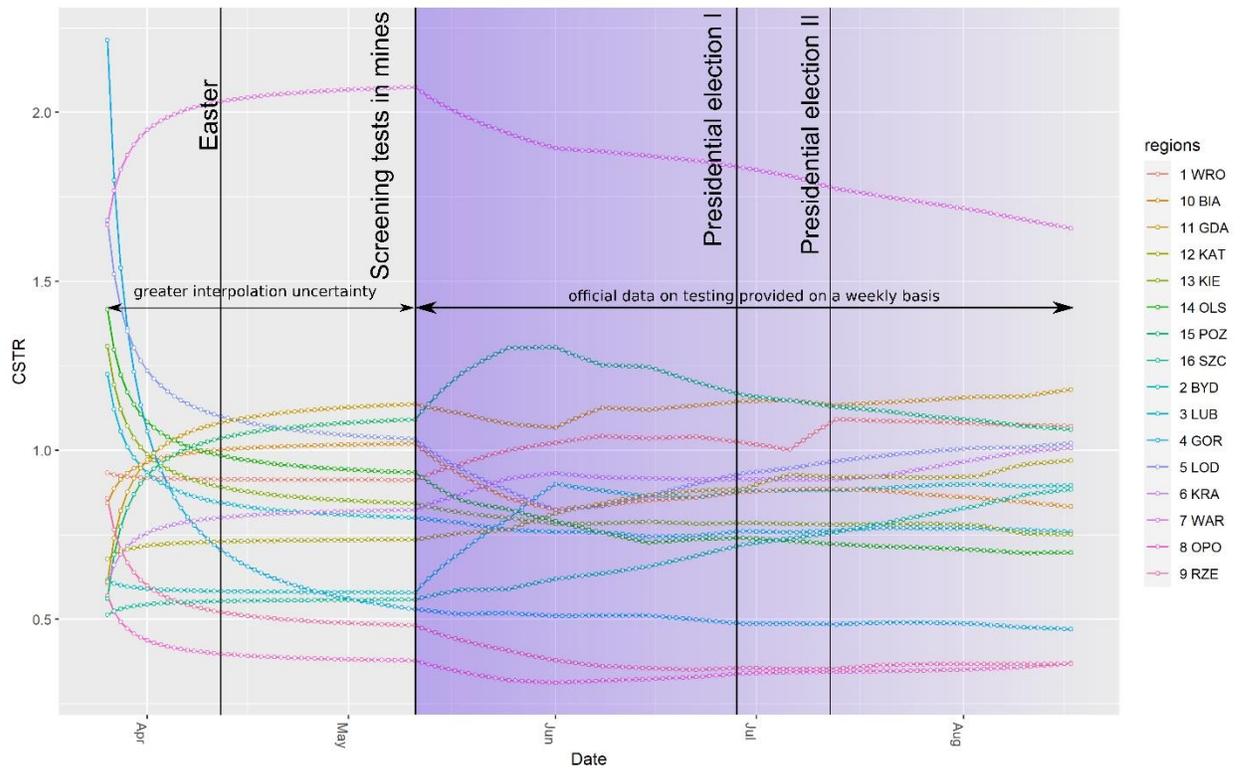

**Fig. S3** Relative testing intensity throughout the study period. Please note that the data are interpolated which causes "edge" effects at the nodes

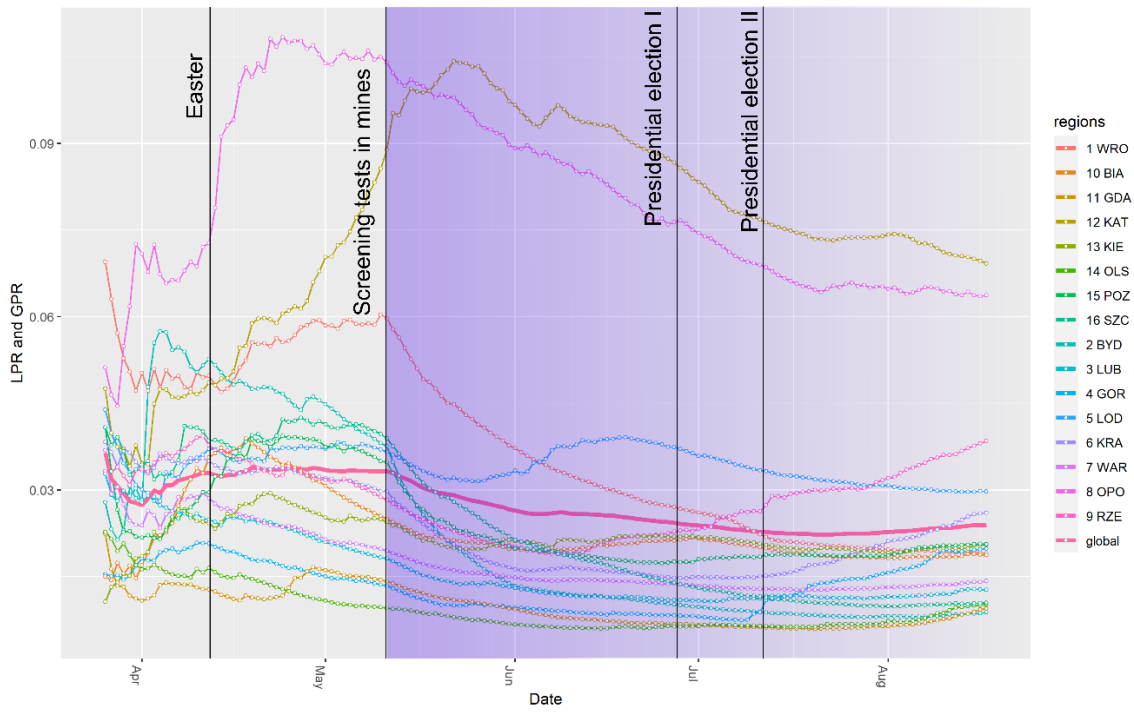

**Fig. S4** Local (LPR) and global (GPR) cumulative positivity ratios

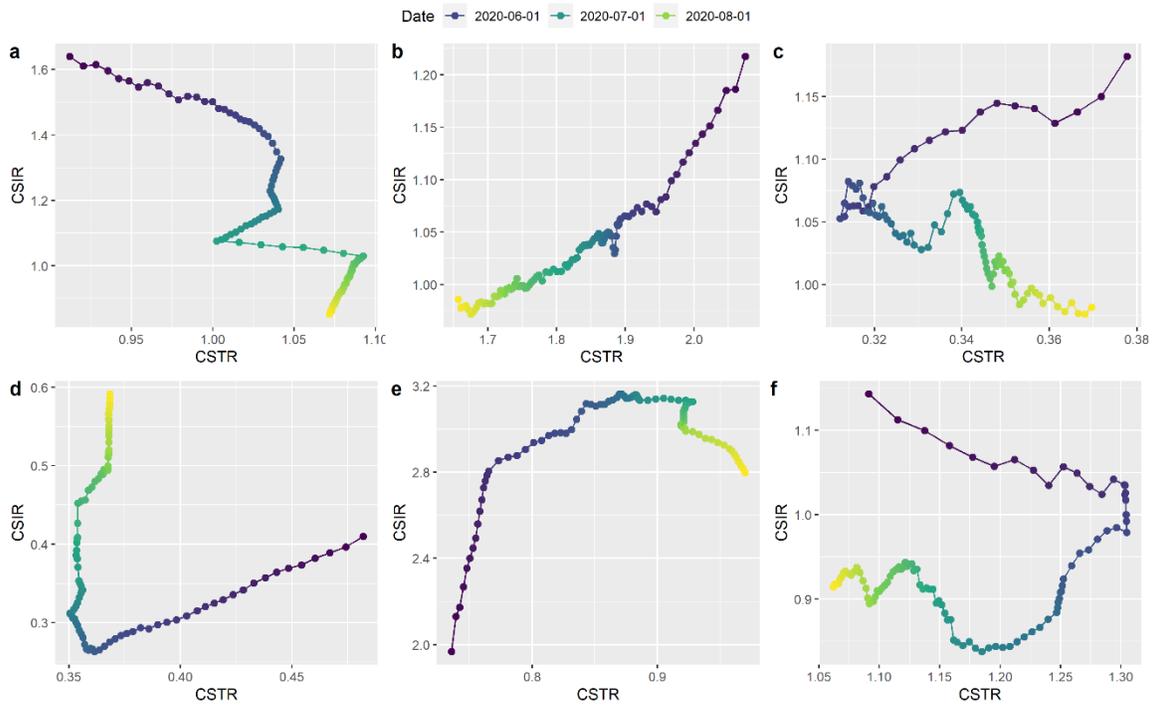

**Fig. S5** Relationship between CSTR and CSIR: (A) 1 WRO: CSIR is monotonically decreasing even if the CSTR is sometimes increasing (B) 7 WAR: both CSTR and CSIR are decreasing up to early August, (C) 8 OPO: decreasing CSTR with a simultaneous growth in CSIR in late-May, (D) 9 RZE: CSIR increasing since mid-June, yet CSTR increased in August, (E) 12 KAT: similar growth rate between CSTR and CSIR in early June that followed a greater growth rate of CSIR in May, (F) 15 POZ: decreasing CSTR with undulating CSIR

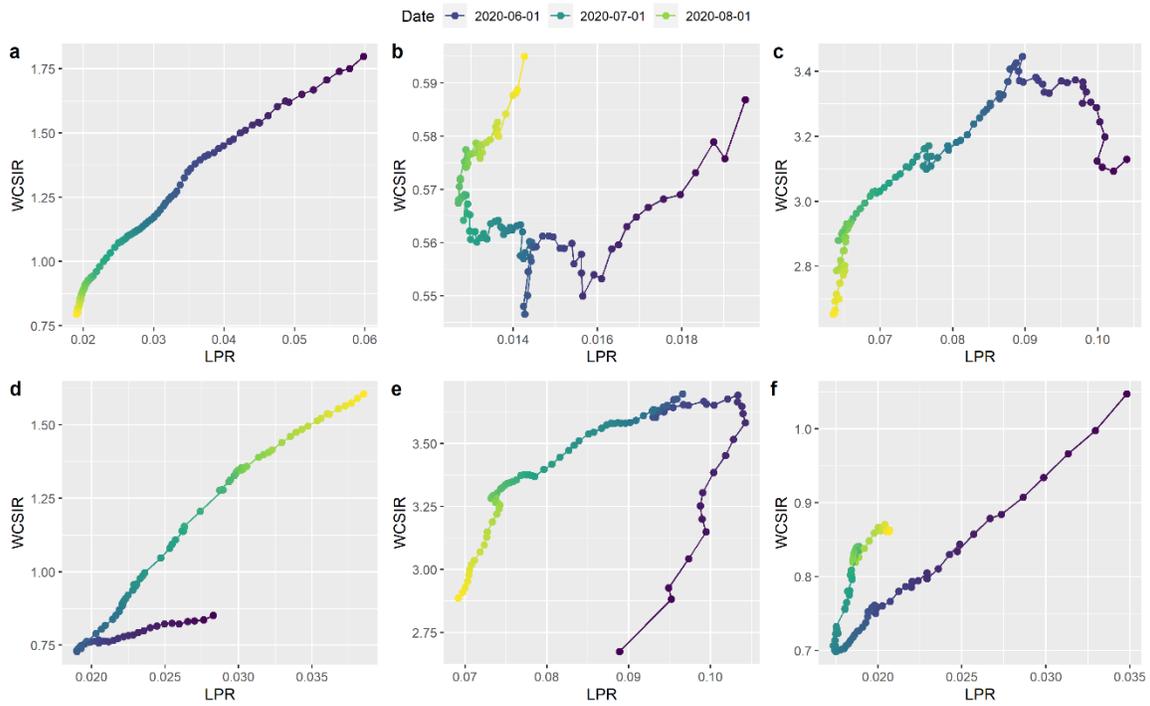

**Fig. S6** Relationship between LPR and WCSIR: (A) 1 WRO: simultaneous decrease in LPR and WCSIR (B) 7 WAR: constant LPR with increasing WCSIR in mid-July, (C) 8 OPO: slightly decreasing LPR with an increasing WCSIR in late-May, (D) 9 RZE: both LPR and WCSIR are increasing since June, (E) 12 KAT: LPR and WCSIR are together increasing and then together decreasing, (F) 15 POZ: faster growth of WCSIR than that of LPR since July

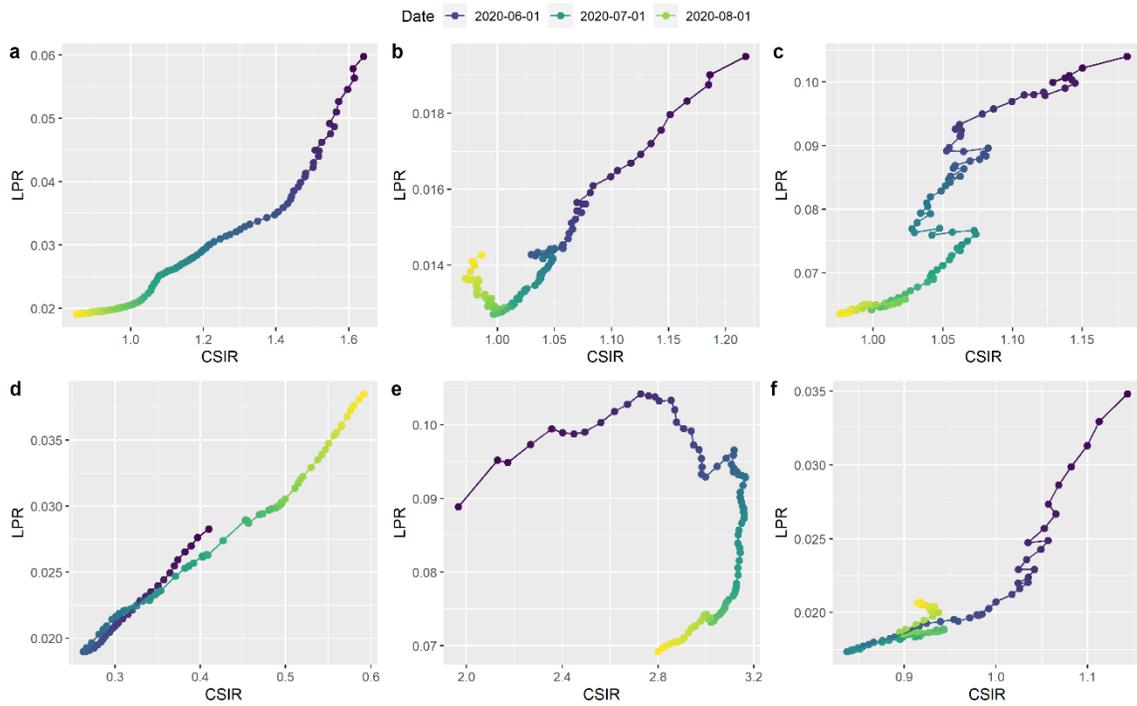

**Fig. S7** Relationship between CSIR and LPR: (A) 1 WRO: both CSIR and LPR are decreasing (B) 7 WAR: CSIR is decreasing and LPR is increasing, (C) 8 OPO: a zigzag trajectory, trend shows decreasing CSIR and LPR, (D) 9 RZE: both CSIR and LPR are increasing since mid-June, (E) 12 KAT: constant CSIR with decreasing LPR in July, (F) 15 POZ: change in simultaneous decrease of both CSIR and LPR since July

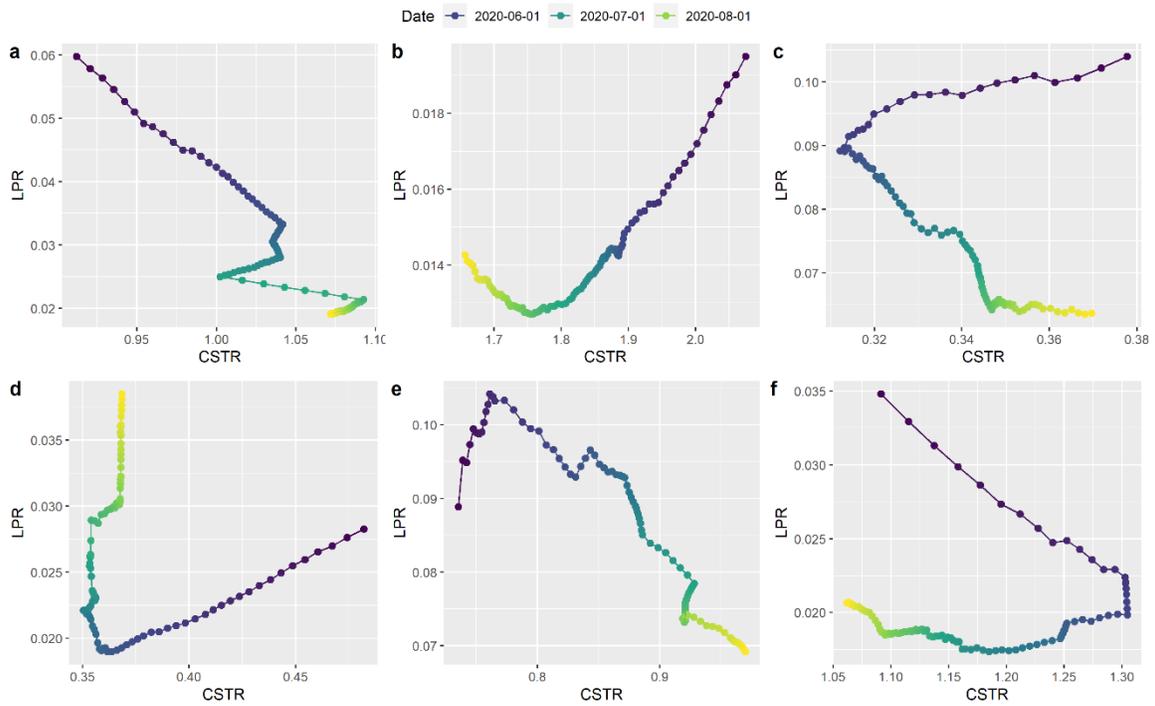

**Fig. S8** Relationship between CSTR and LPR: (A) 1 WRO: LPR is monotonically decreasing, CSTR is sometimes increasing (B) 7 WAR: CSTR is decreasing and LPR increasing since late-July, (C) 8 OPO: CSTR increasing since June with a constant LPR in August, (D) 9 RZE: CSTR increased slightly in August with dramatically increasing LPR, (E) 12 KAT: decreasing LPR and increasing CSTR since mid-June, (F) 15 POZ: faster decrease rate for CSTR than the rate of increase of LPR since July

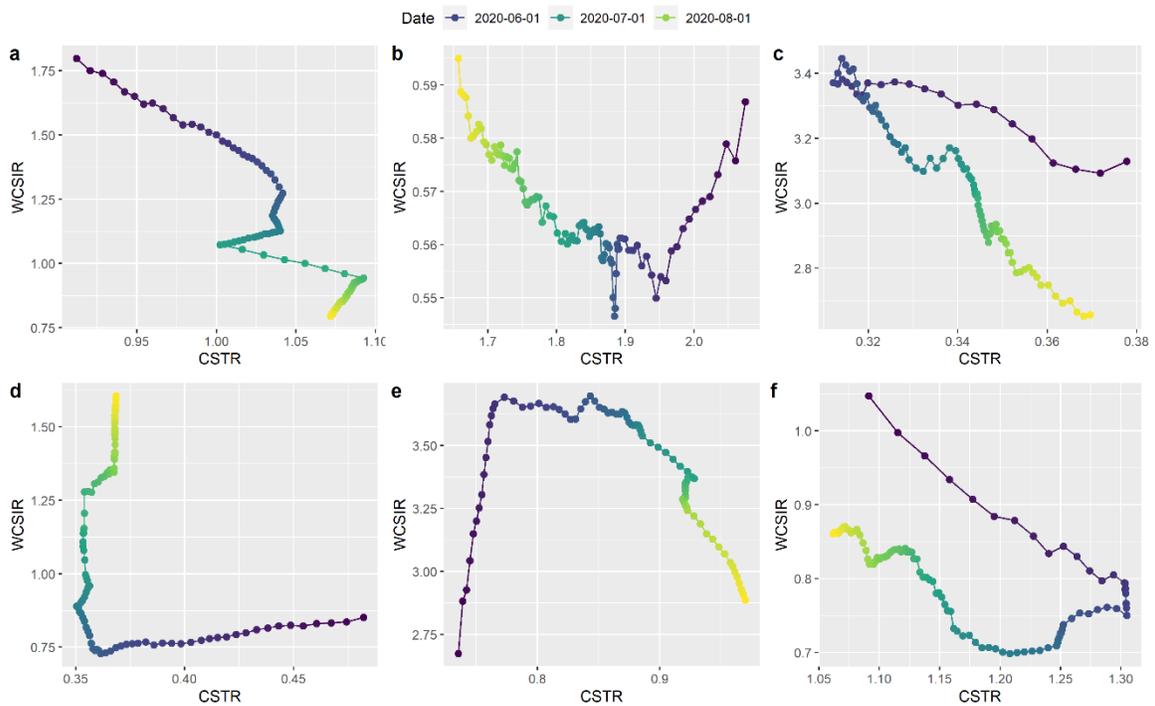

**Fig. S9** Relationship between CSTR and WCSIR: (A) 1 WRO: WCSIR is monotonically decreasing, the CSTR is sometimes increasing (B) 7 WAR: CSTR decreasing with a trend of increasing WCSIR since late-May, (C) 8 OPO: decreasing CSTR with a simultaneous growth in WCSIR in mid-May, (D) 9 RZE: CSTR increased slightly in August with a dramatically increasing WCSIR, (E) 12 KAT: decreasing WCSIR since mid-June and increasing CSTR, (F) 15 POZ: decreasing CSTR with undulating WCSIR since July

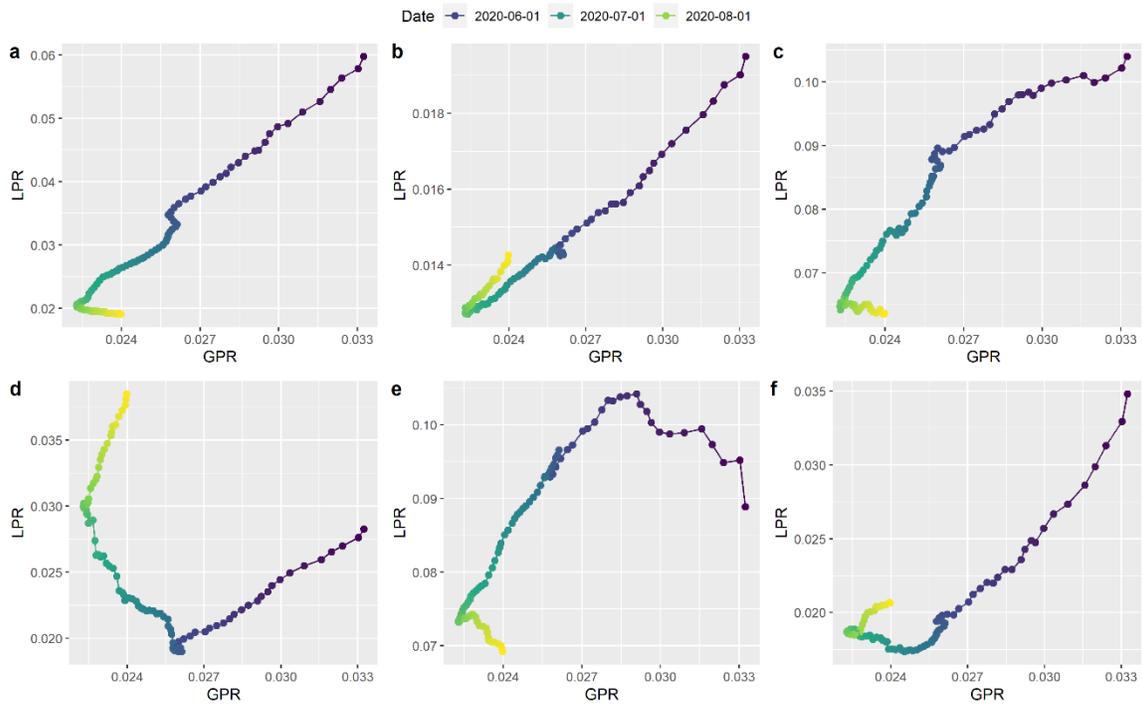

**Fig. S10** Relationship between GPR and LPR: (A) 1 WRO: decreasing LPR with increasing GPR in early-June and August, (B) 7 WAR: in August both LPR and GPR are increasing, (C) 8 OPO: in August LPR approximately constant with increasing GPR, (D) 9 RZE: in August LPR increasing faster than GPR, (E) 12 KAT: increasing LPR with decreasing GPR in May, (F) 15 POZ: in August GPR increasing faster than LPR

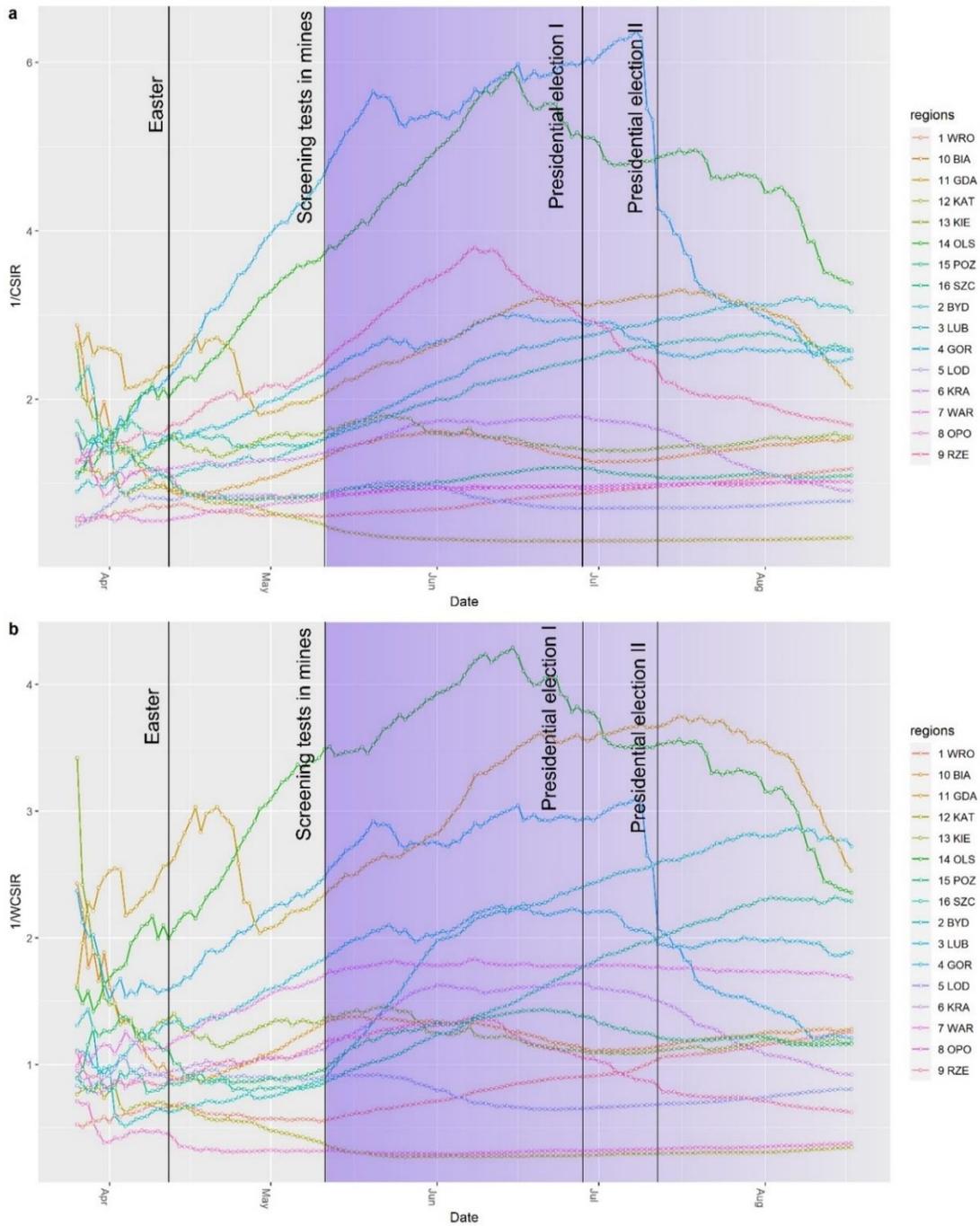

**Fig. S11** Relative safety perspective assumed as 1/risk: (A) unweighted safety (1/CSIR); (B) weighted safety (1/WCSIR)

## Supplementary Tables

**Tab. S1** Reported inaccuracies related to the number of confirmed cases

| Error ID | No. of errors | Reporting date | Error date | Correction date | Reference |
|---|---|---|---|---|---|
| 1 | 4 | June 26, 2020 | ND | June 25, 2020 | [1] |
| 2 | 5 | June 26, 2020 | ND | June 24, 2020 | [2] |
| 3 | 51 | June 19, 2020 | ND | June 16-19, 2020 | [3] |
| 4 | 1 | June 2, 2020 | ND | June 1, 2020 | [4] |
| 5 | 2 | June 9, 2020 | June 6, 2020 | June 6, 2020 | [5] |
| 6 | 4 | May 31, 2020 | ND | May 31, 2020 | [6] |
| 7 | 1 | May 29, 2020 | ND | May 27, 2020 | [7] |
| 8 | 2 | May 26, 2020 | ND | May 26, 2020 | [8] |
| 9 | 1 | June 1, 2020 | ND | May 25, 2020 | [9] |
| 10 | 34 | May 25, 2020 | ND | May 25, 2020 | [10] |
| 11 | 4 | May 23, 2020 | ND | May 23, 2020 | [11] |
| 12 | 2 | May 25, 2020 | ND | May 23, 2020 | [12] |
| 13 | 39 | May 13, 2020 | May 12, 2020 | May 12, 2020 | [13] |
| 14 | 21 | May 8, 2020 | May 7, 2020 | May 7, 2020 | [14] |
| 15 | 17 | May 7, 2020 | May 5, 2020 | May 5, 2020 | [15] |
| 16 | 2 | May 6, 2020 | ND | May 5, 2020 | [16] |
| 17 | 63 | April 30, 2020 | April 16 - 19, 2020 | April 16 - 19, 2020 | [17] |
| 18 | 5 | June 2, 2020 | ND | June 1, 2020 | [18] |

**Tab. S2** Cumulative number of tests, as of March 26, 2020

| Region | Cumulative number of tests (as of March 26, 2020) | Source (OD – official data sent on request) |
|---|---|---|
| 1 WRO | 2360 | OD |
| 2 BYD | 1112 | OD |
| 3 LUB | 2254 | OD |
| 4 GOR | 1953 | OD |
| 5 LOD | 3599 | OD |
| 6 KRA | 1803 | OD |
| 7 WAR | 7890 | OD |
| 8 OPO | 488 | OD |
| 9 RZE | 1591 | [19] |
| 10 BIA | 867 | [19] |
| 11 GDA | 1248 | OD |
| 12 KAT | 2672 | OD |
| 13 KIE | 1408 | OD |
| 14 OLS | 1758 | OD |
| 15 POZ | 1715 | OD |
| 16 SZC | 759 | OD |

**SI References**

12. Ministry of Health Poland. Report on errors #12 [Internet]. 2020. Available from: https://twitter.com/MZ_GOV_PL/status/1264942629856346114

13. Ministry of Health Poland. Report on errors #13 [Internet]. 2020. Available from: https://twitter.com/MZ_GOV_PL/status/1260482119446519810

14. Ministry of Health Poland. Report on errors #14 [Internet]. 2020. Available from: https://twitter.com/MZ_GOV_PL/status/1258781443997732866%0D%0A

15. Ministry of Health Poland. Report on errors #15 [Internet]. 2020. Available from: https://twitter.com/MZ_GOV_PL/status/1258419202605617155

16. Ministry of Health Poland. Report on errors #16 [Internet]. 2020. Available from: https://twitter.com/MZ_GOV_PL/status/1257943323106586626

17. Ministry of Health Poland. Report on errors #17. 2020; Available from: https://twitter.com/MZ_GOV_PL/status/1255883292370448384

18. Ministry of Health Poland. Report on errors #18 [Internet]. 2020. Available from: https://twitter.com/MZ_GOV_PL/status/1267841822455463937

19. Sokalska A. Testy, testy, testy - podkreśla WHO. Ile robi się ich u nas i czy dane są transparentne? Pytamy urzędników o badania na koronawirusa. Pol Times [Internet]. 2020; Available from: https://polskatimes.pl/testy-testy-testy-podkresla-who-ile-robi-sie-ich-u-nas-i-czy-dane-sa-transparentne-pytamy-urzednikow-o-badania-na-koronawirusa/ar/c1-1488489951